\documentclass[%
 reprint,
 amsmath,amssymb,
 aps, prl,
]{revtex4-2}

\usepackage{morefloats}
\usepackage{color}
\usepackage{epsfig,graphicx,amsfonts,amsbsy}
\usepackage{amsmath,amsfonts,amsthm,amssymb}
\usepackage{appendix}
\usepackage{bbm}
\usepackage{makeidx}
\usepackage{url}
\usepackage{verbatim}
\usepackage{mathrsfs} 
\usepackage{morefloats}
\usepackage{appendix}
\usepackage{bbm}
\usepackage{makeidx}
\usepackage{url}
\usepackage{verbatim}
\usepackage[bookmarksnumbered,pdfpagelabels=true,plainpages=false,colorlinks=true,linkcolor=blue,citecolor=blue,urlcolor=blue]{hyperref}
\usepackage[bookmarksnumbered,pdfpagelabels=true,plainpages=false,colorlinks=true,linkcolor=blue,citecolor=blue,urlcolor=blue]{hyperref}
\usepackage[section]{placeins}
\usepackage{array}
\usepackage{booktabs}
\usepackage{multirow}
\usepackage{bbm}
\usepackage{tabularx}
\usepackage{cancel,soul}
\usepackage{nicefrac, xfrac}
\usepackage{ulem}
\usepackage{bbold}
\usepackage{mathtools}
\usepackage{float}
\DeclarePairedDelimiter\bra{\langle}{\rvert}
\DeclarePairedDelimiter\ket{\lvert}{\rangle}
\DeclarePairedDelimiterX\braket[2]{\langle}{\rangle}{#1\,\delimsize\vert\,\mathopen{}#2}

\newcommand{\balpha}{\boldsymbol{{\rm \alpha}}}
\newcommand{\sectioncustom}[1]{\noindent {\it #1.- }}

\def \fcfm {Departamento de F{\'i}sica, Facultad de Ciencias Físicas y Matemáticas, CEDENNA, Universidad de Chile, Santiago, Chile.}
\def \jgu {Institute of Physics, Johannes Gutenberg University Mainz, 55099 Mainz, Germany.}
\def \uta {Departamento de Física, Facultad de Ciencias, Universidad de Tarapacá, Casilla 7-D, Arica, Chile.}

\usepackage[utf8]{inputenc}
\begin{document}
\title{Antiferromagnetic Pure Spin Current Memdevices}
\author{Martin Latorre$^1$}
\author{Gaspar De la Barrera$^2$}
\author{Roberto E. Troncoso$^3$}
\author{Alvaro S. Nunez$^1$} 
\affiliation{$^1$\fcfm}
\affiliation{$^2$\jgu}
\affiliation{$^3$\uta}
\date{\today}

\begin{abstract}
Spin currents can be generated through various mechanisms, including the piezospintronic effect, which arises when strain or lattice distortions induce a change in the dipolar spin moment, causing a pure spin current without necessarily being accompanied by net charge transport. This opens new possibilities for low-power information processing and novel device architectures. In this work, we propose a novel effect, the spintronic-magneto-impedictive effect, as the theoretical basis for a pure spin-current memory-like device based on antiferromagnetic components. We focus on materials that can be modeled by the so-called spin-Rice-Mele Hamiltonian, incorporating a magnetic field gradient that explicitly breaks inversion symmetry. Our results shed light on how spin currents are generated and controlled, providing new insights into the potential of these materials for next-generation spintronic technologies.
\end{abstract}
\maketitle

{\it Introduction.--}  Spin-currents are a central concept in spintronics \cite{Wang2025, Dey2021, Maekawa2017}, describing the transport of spin angular momentum through a material. In contrast to conventional charge currents, spin-currents are even under time reversal \cite{Maekawa2023} and can propagate without an accompanying flow of net charge, thereby enabling fundamentally new paradigms for low-power information processing and spin-based device architectures. Spin-currents can be generated via spin polarized charge injection from ferromagnets \cite{Johnson2017}, spin Hall and inverse spin Hall effects in materials with strong spin-orbit coupling \cite{Sinova2015}, and dynamical processes such as spin-pumping \cite{Takahashi2016} and the spin Seebeck effect \cite{Uchida2008}. They can be manipulated and detected through spin-orbit torques \cite{Shao2021}, nonlocal spin-valve geometries \cite{Ji2007}, and optical probes \cite{vanDriel2006}.
Beyond applications, spin-currents provide a versatile platform for exploring fundamental condensed-matter phenomena \cite{Desmarais2024}. Their coupling to magnetization dynamics, quantum coherence, and topological electronic states continues to drive progress across a wide range of material systems, including transition-metal ferromagnets \cite{Amin2019}, heavy metals \cite{Liu2024}, topological insulators \cite{Zhang2016}, and antiferromagnets \cite{Takahashi2008}.


Antiferromagnetic materials provide an ideal platform for spin transport due to their combination of vanishing net magnetization and robust spin dynamics \cite{Baltz2018, Jungwirth2016, DalDin2024}. Unlike ferromagnets, their compensated magnetic order suppresses stray fields and enables ultrafast dynamics in the terahertz regime. Spin currents in antiferromagnets (AFs) can be carried by itinerant electrons through spin-dependent scattering or by collective excitations such as magnons. Importantly, their symmetry allows efficient generation and detection of spin-currents via spin–orbit effects while avoiding key limitations of ferromagnetic systems, including cross-talk and magnetic-field sensitivity. Additionally, AFs exhibit long spin diffusion lengths \cite{Lebrun2018,Das2022} and support spin transport in both collinear and noncollinear magnetic textures, making antiferromagnetic spintronics a rapidly advancing field with applications ranging from high-speed memory and logic to energy-efficient spin-based interconnects.

\begin{figure}[h]
\centering
\includegraphics[width=\linewidth]{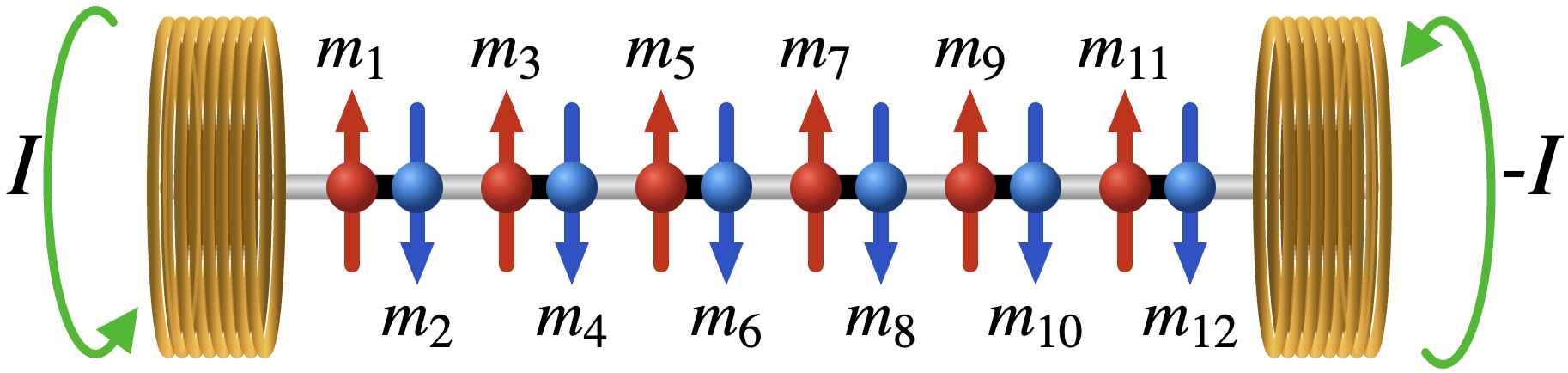}
\caption{Schematic setup for AF Rice-Mele spin chain subjected to a magnetic field gradient. Black and gray bonds represent the short and long links, respectively.}
\label{fig1:AF-chain-memdevice}
\end{figure}

Recent studies indicate that certain AFs exhibit a spin-dependent phenomenon known as the piezospintronic effect \cite{Nunez2014, Ulloa2017, Liu2019, Saez2024, Chen2025, Castro2025}, in which spin-currents are generated and controlled via mechanical deformation in systems with coupled spin, lattice, and electronic degrees of freedom. Analogous to the piezoelectric effect-where strain induces a change in electric polarization-the piezospintronic effect arises when strain or lattice distortions produce a dipolar spin moment, generating a pure spin current without charge flow. This direct coupling enables mechanical control of spin transport and spin accumulation, providing an alternative route to spin-current generation beyond conventional electrical or thermal methods. The effect is particularly pronounced in low-dimensional magnetic insulators, where nanoscale elastic deformations can be engineered. By combining mechanical tunability with low-power operation, the piezospintronic effect opens new avenues for flexible and multifunctional spintronic devices \cite{Saez2024}.

In this Letter, we propose a theoretical framework for a pure spin-current, memory-like device based on antiferromagnetic components \cite{Chua1971, Chua2019, Mazumder2012, Adamatzky2013, Tetzlaff2013}. The central prediction is the spintronic-magneto-impedictive effect, an impedance-like relation between spin-current and an external magnetic-field gradient, as illustrated schematically in Fig. \ref{fig1:AF-chain-memdevice}. Our results advance the understanding of spin-current generation and control, shed light on the piezospintronic effect, and reveal new opportunities for harnessing spin-currents in next-generation spintronic technologies \cite{Sort2025}.

{\it Spintronic magneto-memimpedance.--} The basic system that we will address requires a $\mathcal{PT}$-invariant piezospintronic antiferromagnet such as the one described in \cite{Nunez2014}. We will argue that, when exposed to a magnetic field gradient, the system responds with the emission of a spin current, obeying a linear response relation,
\begin{align}
\boldsymbol{{\rm J}}_S(\omega)=Z_S(\boldsymbol{\alpha}, \omega) \nabla B(\omega),
\end{align}
in frequency space, where $Z_S$ is a pure spin-current magneto-memimpedance \cite{Wakrim2016, MacVittie2013} and  $\boldsymbol{\alpha}$ stands for physical parameters specifying the state of the system, such as the Néel vector or elastic strain. As it turns out  $\boldsymbol{\alpha}$ has a time evolution dictated by ${d\boldsymbol{\alpha}}/{dt}=\mathcal{F}(\balpha, \nabla B)$ with known $\mathcal{F}$. The last two equations are reminiscent of those governing the behavior of a memristor \cite{Chua2019}. Control of magnetic excitations via a magnetic field gradient has been explored in several works in the context of magnetic straintronics \cite{VidalSilva2020, Pal2023, Majumder2023, Bandyopadhyay2022}. 
Coupling to a magnetic field gradient entails three distinct effects. Fisrt, we have the action of $\nabla B$ directly on the dynamics of $\mathbf{n}$ as a consequence of the Zeeman effect. Second, the gradient modifies the electronic spectrum, explicitly breaking inversion symmetry. Finally, it exerts opposite forces on the magnetic moments of the two antiferromagnetic sublattices, inducing a lattice distortion that enhances dimerization.

The memory in this device is stored in the time-dependent state of its Néel vector $\mathbf{n}$ and in the lattice distortion $u$. In this analogy, the read aspects of the memory are provided by the generated spin current.  In a generic system, there are three contributions to $Z_S(\boldsymbol{\alpha},\omega)$: 
the piezospintronic response $\lambda_{\rm pzsp}={\partial\mathbf{P}_S}/{\partial u}$ 
\cite{Nunez2014}, the Néel-vector modulation 
$\lambda_{n}={\partial\mathbf{P}_S}/{\partial n}$, and the 
magnetization-gradient contribution 
$\lambda_{\nabla B}={\partial\mathbf{P}_S}/{\partial (\nabla B)}$. They can be cast in the equation,
\begin{equation}\label{eq:js_full}
\mathbf{J}_S=\frac{d\mathbf{P}_S}{dt}=\lambda_{\rm pzsp} \frac{d{\bf u}}{dt}+\lambda_{n} \frac{d\mathbf{n}}{dt}+\lambda_{\rm\nabla B}\frac{d\mathbf{\nabla B}}{dt}.
\end{equation}
Modeling the mechanical degrees of freedom as an overdamped particle subjected to the force $g\mu_B\nabla B$ \cite{Jackson1998, Griffiths2023, Zangwill2012}, we obtain:
\begin{align}
Z_S(\boldsymbol{n},u,\omega)=(\mu\,\lambda_{\rm pzsp}+\nu\,\lambda_{n})+i\,\omega\,\zeta\,\lambda_{\rm\nabla B},
\end{align}
where the parameters $\mu$, $\nu$, and $\zeta$, depend on the mechanical and magnetic states of the system.

{\it Microscopic model.--} We now focus on certain materials that can be modeled by the so-called spin-Rice-Mele Hamiltonian \cite{Saez2024, Vergara2024, Saez2023}. Within this class of materials, we can find systems such as one-dimensional MoX$_3$ (X = Cl, Br, I) \cite{Mella2024} and FeOOH \cite{Chen2025}.
The Hamiltonian for the system can be written:
 \begin{align}
\mathcal{H}_{\rm el} = 
- \sum_{n} &\nonumber\left[ {\rm t}_n(\mathbf{u}) \, 
c_{n+1,\sigma}^\dagger c_{n,\sigma} +\mbox{h.c.}\right.\\
&\qquad\quad\left.+ (-1)^nc_{n,\sigma}^\dagger \left(\Delta\mathbf{n} \cdot\boldsymbol{\tau}_{\sigma\sigma^\prime}\right)c_{n,\sigma^\prime}  \right],
\end{align}
where $\mathbf{n}$ stands for the Néel vector of the antiferromagnetic chain. The dimerized hopping amplitude is ${\rm t}_n(\mathbf{u})={\rm t}+(-1)^n\delta {\rm t}(\mathbf{u})$, being the parameter $\delta {\rm t}(\mathbf{u})$ dependent on the lattice deformation. The spin polarization depends on time through ${\bf u}$ and $\mathbf{n}$. These, in turn, behave according to the classical overdamped equations \cite{Landau1976, Baltz2018, Yuan2021},
\begin{align}
\gamma\frac{d u}{dt} + k(u - u_0) + g\mu_B\,{n_z}\nabla B &= 0
\label{eq:u_eom}
,\\
\mathbf{n}\times\left(m\frac{d^2\mathbf{n}}{dt^2} + \alpha_n\frac{d\mathbf{n}}{dt} 
- a_n\left(\mathbf{h}_{\mathbf{n}} + g\mu_B(\nabla B)\hat{\bf z}\right)\right) &= 0,
\label{eq:n_eom}
\end{align}
where $k$ is the elastic constant of the material, $\gamma$ is the 
damping coefficient of the lattice displacement, $u_0$ represents 
the dimerization at equilibrium, $a_n$ is the homogeneous exchange 
coefficient, and $h_{n}$ is the uniaxial 
anisotropy constant. The parameter 
$m = a_n h_{n}/\omega_{\rm AFMR}^2$ represents the effective 
inertia associated with the Néel order parameter, determined by 
$a_n$, $h_{n}$, and the antiferromagnetic resonance frequency 
$\omega_{\rm AFMR}$, while $\alpha_n = \sqrt{a_n h_n}/Q$ is the 
Néel damping parameter, characterizing the dissipation of the 
staggered order dynamics, with $Q=10$ the quality factor of the 
antiferromagnetic resonance, and $\mathbf{h}_{\mathbf{n}}$ the 
uniaxial anisotropy field. A magnetic-field gradient enters as a perturbation directed along 
the $z$-axis, $\nabla B\,\hat{\bf z}$, where $\nabla B = \partial_z B$, and contributes through a position-dependent Zeeman coupling. This term can be decomposed into (i) a contribution proportional to the spatial variation of the eigenstates and (ii) a sublattice-resolved coupling to the local spin moments. The resulting perturbation reads,
\begin{align}\label{eq:H_grad}
    \mathcal{H}_{\nabla B}=-\,g\mu_B\,(\nabla B)\sum_{n}\Big[x_{nA}\,S_{nA}^{z}+x_{nB}\,S_{nB}^{z}\Big],
\end{align}
where $g\mu_B$ is the effective magnetic moment, and the intra-unit-cell displacement $\beta$ defines the sublattice positions $x_{nA}=na-\beta$ and $x_{nB}=na+\beta$. The spin operator is given by $S_{ns}^{z}=(c_{n,s}^{\dagger}\sigma_z c_{n,s})/2$, with $s\in{A,B}$ and $\sigma_z$ the $z-$component of the Pauli matrix vector. This perturbation provides a controlled, adiabatic modulation of the dimerization.
\begin{figure*}[!htb]
    \centering
    \includegraphics[width=\linewidth]{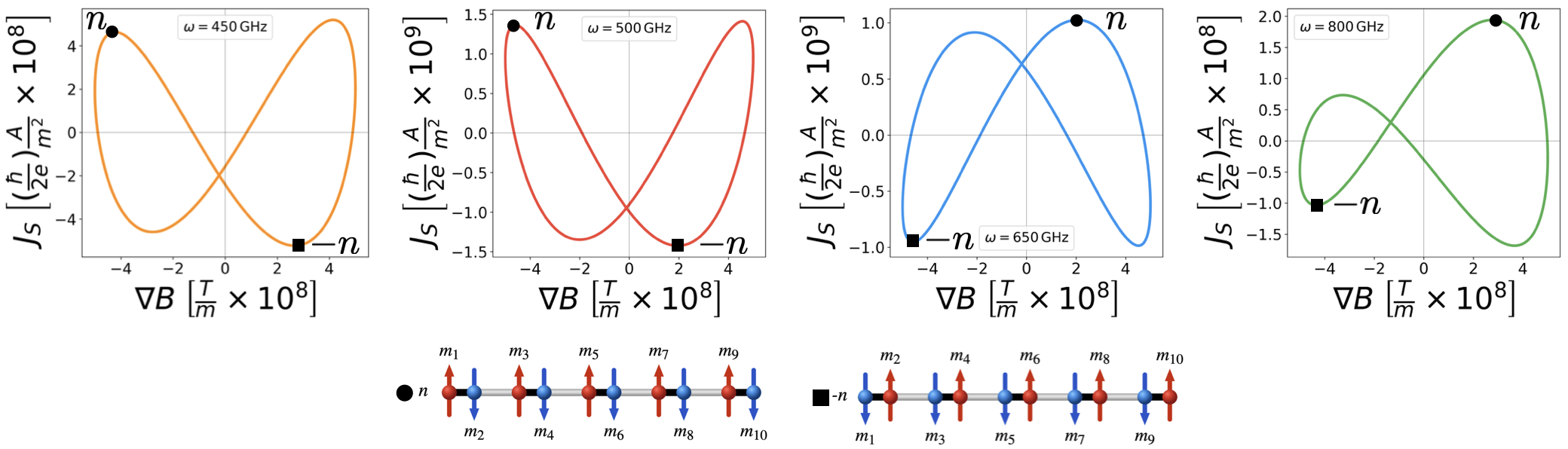}
    \caption{Simulated spin-current curves as a function of the magnitude of the magnetic-field gradient applied to FeOOH chains, for driving frequencies of 450, 500, 650, and 800 $\left[\mathrm{GHz}\right]$ in panels (a)--(d), respectively. The magnetic-field gradient is of the order of $\nabla B = 10^{8}\,\left[\mathrm{T/m}\right]$, which is within the experimentally relevant range \cite{Zablotskii2016-ww}.}
    \label{fig:hysteresisgraph2}
\end{figure*}

{\it Results.--} By analyzing the perturbative contribution $\mathcal{H}_{\nabla B}$, it is necessary to treat the system response within the framework of degenerate perturbation theory, thereby obtaining the eigenvectors associated with this effect. One way to verify the existence of a significant effect arising from 
the inclusion of this term is to compute the spin polarization within 
linear response; for this purpose, we use:
$$
\mathbf{P}_{S} = \sum_{\sigma } \frac{a}{2\pi} \int_{BZ} dk \,i
\bra{\psi_{-,\sigma}} \vec{\tau} \vert\partial_{k} \psi_{-,\sigma} \rangle 
$$
where $\ket{\psi_{-,\sigma}} = \ket{\psi^{(0)}_{-,\sigma}} + \nabla B\,\ket{\psi_{-,\sigma}^{(1)}}$,
$\sigma \in \{\uparrow,\downarrow\}$ denotes the electron spin, and $\vec{\tau} = \vec{\sigma} \otimes \mathbb{1}$ 
is the sublattice operator with eigenvalues $\pm 1$ corresponding to sublattices A and B. In the antiferromagnetic ground state, the Néel order couples spin  to sublattice occupation, such that the contributions from spin-$\uparrow$ and spin-$\downarrow$ channels to $P_S$ are finite and of opposite sign, yielding a net spin-resolved polarization. To first order in the gradient $\nabla B$, we obtain the modification of the spin polarization, $\mathbf{P}_S=\mathbf{P}_S^{(0)}+\nabla B\; \mathbf{P}_S^{(1)}$. Here, $\mathbf{P}_S^{(0)}$ denotes the intrinsic spin polarization of the Rice–Mele lattice, while $\mathbf{P}S^{(1)}$ arises from the first-order correction to the eigenstates $\ket{\psi{-,\sigma}}$.
It is worth noting that, $\mathbf{P}_S$ depends on $u$, $\mathbf{n}$, and $\nabla B$. Consequently, the polarization current in Eq. $\ref{eq:js_full}$ can be expanded to first order in $\nabla B$, with coefficients $\lambda$ quantifying the contribution of each system variable to the polarization current. Their explicit form reads,
\begin{align}
\lambda_{\rm pzsp}
&= \frac{\partial (\delta t)}{\partial u}
   \frac{a}{2\pi} \sum_{\sigma}\int_{-\pi}^\pi 
   \Omega^{\tau^z}_{\delta t,\, k}\,dk
\label{eq:lambda_pzsp}
\\
\lambda_{n}
&= \frac{a}{2\pi} \sum_{\sigma}\int_{-\pi}^\pi 
   \Omega^{\tau^z}_{n,\, k}\,dk
\label{eq:lambda_n}
\\
\lambda_{\nabla B}
&= \frac{a}{2\pi} \sum_{\sigma}\int_{-\pi}^\pi
   \Omega^{\tau^z}_{\nabla B,\, k}\,dk
\label{eq:lambda_gradB}
\end{align}
where $\Omega^{\tau^z}_{x,k} \equiv i\left(\bra{\partial_x\psi_{-,\sigma}}
\tau^z\ket{\partial_k\psi_{-,\sigma}}-\bra{\partial_k\psi_{-,\sigma}}
\tau^z\ket{\partial_x\psi_{-,\sigma}}\right)$ is a generalized Berry 
curvature with respect to parameter $x$ and crystal momentum $k$.

From Eq.~(\ref{eq:n_eom}), the magnetic-field gradient $\nabla B\,\hat{\bf z}$ 
exerts a torque on $\mathbf{n}$ with two components: one lying in the 
plane spanned by $\mathbf{n}$ and $\hat{\bf z}$, which couples directly 
to the anisotropy field $\mathbf{h}_{\mathbf{n}}$ and governs the 
dominant dynamics, and one perpendicular to this plane, which is of 
higher order in the oscillation amplitude and can therefore be neglected. 
This reduces the problem to an effective one-dimensional dynamics within 
the plane containing $\mathbf{n}$ and $\hat{\bf z}$.

\begin{figure}[!htb]
    \centering
    \includegraphics[width=1\linewidth]{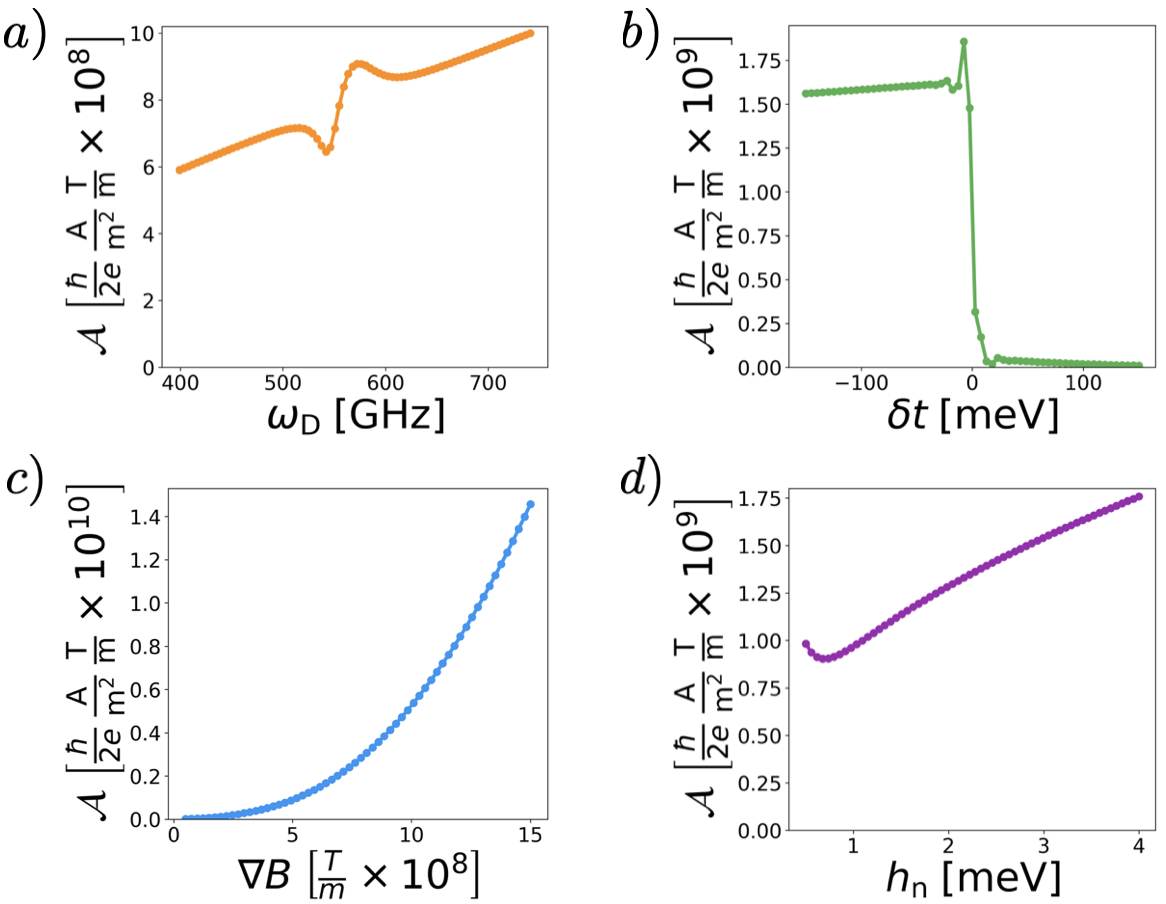}
    \caption{Plots of the system's memristive effect $\mathcal{A}$ as a function of different control parameters: 
(a) $\mathcal{A}$ versus the driving frequency $\omega_D$ associated with the magnetic-field gradient; 
(b) $\mathcal{A}$ versus the dimerization $\delta t$; 
(c) $\mathcal{A}$ versus the magnitude of the magnetic-field gradient $\nabla B$; and 
(d) $\mathcal{A}$ versus the anisotropy parameter $h_n$.}
    \label{memristor_effect}
\end{figure}

To better understand the behavior of the spin current $\mathbf{J}_{S}$, it is important to study the different contributions. Each contribution is governed by its corresponding $\lambda$-parameter, allowing us to distinguish the effect of $\nabla B$ on $u$, on $\mathbf{n}$, and the intrinsic contribution of the gradient itself. After identifying and explicitly calculating the different contributions to the spin current, their combined effect can be analyzed together with the classical overdamped equations of motion for $u$ and $\mathbf{n}$, thereby determining the full behavior of $\mathbf{J}_{S}$ as a function of the applied external magnetic-field gradient. This dependence is presented in Fig.~\ref{fig:hysteresisgraph2}, where several curves are shown for different values of the driving frequency $\omega$ associated with the oscillating magnetic-field gradient, considering regimes both below and above the antiferromagnetic resonance frequency, $\omega_{\mathrm{AFMR}} \approx 570~\mathrm{GHz}$, of FeOOH \cite{Chou2017-cq}.
In this way, it becomes evident that the spin-current response 
exhibits hysteretic cycles, thereby giving rise to purely magnetic memristors governed by the effect of the magnetic-field gradient on the spins. To achieve a more complete understanding of these memristive effects, the memristive response, $\mathcal{A}=\oint J_S\, d\nabla B$, was studied as a function of different control parameters of the system under consideration, leading to the results shown in Fig.~\ref{memristor_effect}.

{\it Discussion.--} We now present the results of our analysis of the spintronic-magneto-impedictive effect in antiferromagnetic spin chains described by the spin-Rice-Mele Hamiltonian. The central quantity of interest is the spin current $J_S$ as a function of the applied magnetic-field
gradient $\nabla B$, which enters the system through three distinct channels: the
piezospintronic coupling $\lambda_\text{pzsp}$, the N\'{e}el vector coupling $\lambda_n$,
and the intrinsic gradient coupling $\lambda_{\nabla B}$. Together, these three
contributions govern the full memristive response of the system.

\underline{Hysteretic spin-current loops.}---Figure~\ref{fig:hysteresisgraph2} displays the simulated
spin-current curves as a function of the magnitude of $\nabla B$ for FeOOH chains,
computed at driving frequencies of 450, 500, 650, and 800~GHz, spanning regimes both
below and above the antiferromagnetic resonance frequency
$\omega_\text{AFMR} \approx 570$~GHz of FeOOH~\cite{Chou2017-cq}. The most salient feature
of these curves is the appearance of \textit{hysteretic loops} in all frequency regimes,
directly establishing the memristive character of the spin-current response. These loops
are a direct consequence of the time-dependent state variables---the N\'{e}el vector
$\mathbf{n}$ and the lattice distortion $u$---which evolve according to their respective
overdamped equations of motion [Eqs.~(\ref{eq:u_eom}) and~(\ref{eq:n_eom})] and retain
information about the history of the applied field gradient. The system therefore satisfies
the defining characteristic of a memristor: its instantaneous response depends not only on
the present drive, but on the integrated past dynamics of the internal state.
 
\underline{Frequency dependence across} $\omega_\text{AFMR}$.---The shape and area of the
hysteretic loops depend sensitively on the driving frequency $\omega$. Below resonance
[panels~(a)--(b), $\omega = 450$ and $500$~GHz], the loops are wide and display a
pronounced asymmetry, reflecting the dominant role of the piezospintronic and
N\'{e}el-vector contributions to the spin current. As the frequency crosses
$\omega_\text{AFMR}$ and enters the above-resonance regime [panels~(c)--(d),
$\omega = 650$ and $800$~GHz], the loop morphology changes qualitatively: the curves
narrow and the overall magnitude of $J_S$ is reduced, consistent with the overdamped
dynamics of $\mathbf{n}$ becoming less responsive to the oscillating gradient at high
frequencies. At $800$~GHz the loop recovers a larger amplitude, which we attribute to an
interplay between the inertial term in the N\'{e}el-vector equation of motion and the
intrinsic $\lambda_{\nabla B}$ contribution, which becomes dominant when the mechanical
degree of freedom $u$ can no longer adiabatically follow the drive. These results confirm
that the resonance frequency of the antiferromagnet serves as a natural threshold
separating qualitatively different memristive regimes.
 
\underline{Memristive response.}---To quantify the strength of the memory effect, we compute
the memristive response $\mathcal{A} = \oint J_S\, d\nabla B$, defined as the area
enclosed by the hysteretic loop in the $J_S$--$\nabla B$ plane. A nonzero $\mathcal{A}$
is the direct signature of memristive behavior, and its dependence on system parameters
is presented in Fig.~\ref{memristor_effect}. Panel~(a) shows that $\mathcal{A}$ peaks sharply near
$\omega_\text{AFMR}$, demonstrating that the memory effect is resonantly enhanced and
that the antiferromagnetic resonance frequency constitutes an optimal operating point for
the device. The peak is relatively narrow in frequency, suggesting that these devices could
additionally function as frequency-selective spin-current switches.
 
\underline{Dependence on dimerization.}---Panel~(b) of Fig.~\ref{memristor_effect} examines the
dependence of $\mathcal{A}$ on the dimerization parameter $\delta t$. The memristive
response increases monotonically with $\delta t$, saturating at large dimerization.
This behavior is consistent with the analytical expression for $\lambda_\text{pzsp}$ in
Eq.~(\ref{eq:lambda_pzsp}): a larger bond alternation amplifies the Berry-phase-like
integral over the Brillouin zone, yielding a stronger piezospintronic coupling and
thus a larger spin polarization response to lattice deformation. This result confirms
that dimerization is a key tuning parameter for optimizing device performance, and that
materials with a pronounced Peierls-type distortion are the most promising candidates.
 
\underline{Dependence on the field gradient amplitude.}---Panel~(c) shows the dependence of
$\mathcal{A}$ on the magnitude of the applied field gradient $\nabla B$. For small
gradients the response grows approximately quadratically, as expected from a perturbative
treatment in which the leading correction to the loop area is second order in the driving
amplitude. At larger $\nabla B$ the response saturates and eventually decreases, signaling
the onset of nonlinear effects beyond the linear-response regime assumed in the derivation
of Eq.~(\ref{eq:js_full}). The characteristic gradient scale at which this crossover
occurs is of order $10^8$~T/m, consistent with values reported in experimental studies of
high-gradient magnetic fields at the nanoscale~\cite{Zablotskii2016-ww}, validating the
physical relevance of the parameter range used throughout our simulations.
 
\underline{Dependence on magnetic anisotropy.}---Panel~(d) illustrates the effect of the
magnetic anisotropy parameter $h_n$ on the memristive response. Increasing $h_n$ stiffens
the N\'{e}el vector dynamics, suppressing large-amplitude fluctuations and reducing the
hysteretic loop area. In the limit of very large anisotropy, $\mathbf{n}$ becomes
essentially frozen along the easy axis and the memory effect is quenched. Conversely,
small anisotropy permits freer precession of the N\'{e}el vector, enriching the internal
dynamics and enhancing $\mathcal{A}$. These results highlight a direct trade-off between
magnetic stability---desirable for retention of the stored state---and the amplitude of
the spin-current response during operation, a competition analogous to the
stability-plasticity dilemma encountered in neuromorphic computing.

{\it Conclusions.--} The picture that emerges from these results is the following. The applied magnetic-field gradient $\nabla B$ simultaneously perturbs the N\'{e}el vector, distorts the lattice through the differential Zeeman force on the two
sublattices, and directly modifies the electronic eigenstates via the
symmetry-breaking term $H_{\nabla B}$ in Eq.~(\ref{eq:H_grad}). Each channel contributes to the spin polarization through the corresponding $\lambda$ coefficient computed in
Eqs.~(\ref{eq:lambda_pzsp})--(\ref{eq:lambda_gradB}). Because $\mathbf{n}$ and $u$ evolve
on their own timescales governed by Eqs.~(\ref{eq:u_eom})--(\ref{eq:n_eom}), the system retains a phase lag with respect to the drive, and the combined state $(\mathbf{n}, u)$ traces a closed orbit in state space that does not coincide on the forward and return sweeps---the hallmark of memristive behavior. The spin-current memimpedance
$Z_S({\boldsymbol{\alpha}},\omega)$ therefore encodes both the instantaneous electronic response and the
cumulative history of the magnetic and mechanical state, realizing a purely spin-based
analogue of the memristor originally conceived by Chua~\cite{Chua1971}.
 
\underline{Comparison with charge-current memristors.}---It is worth contrasting the present
device with conventional charge-current memristors. In the latter, hysteresis originates
from resistive switching driven by ionic migration, phase transitions, or conducting-filament
formation, all of which involve substantial Joule dissipation and electromigration damage
over repeated cycles~\cite{Mazumder2012}. In our antiferromagnetic spin-current memdevice,
no net charge flows and the relevant degrees of freedom---spin texture and lattice
dimerization---are intrinsically reversible and energy-efficient. The absence of stray
fields, characteristic of antiferromagnets, further suppresses cross-talk between
neighboring devices, a critical advantage for high-density integration~\cite{Jungwirth2016}.
These features position the proposed memdevice as a compelling building block for
neuromorphic and unconventional computing architectures that seek to exploit spin degrees
of freedom.
 
\underline{Materials outlook.}---Although our analysis focused on FeOOH as a concrete example,
the theoretical framework applies to any $\mathcal{PT}$-invariant piezospintronic
antiferromagnet described by the spin-Rice-Mele Hamiltonian, including the
one-dimensional MoX$_3$ (X~=~Cl, Br, I) family~\cite{Mella2024}. The tunability of the
resonance frequency, dimerization strength, and magnetic anisotropy across this materials
class offers broad flexibility for targeting specific operational frequency windows and
for optimizing the memory response to given device requirements. We anticipate that recent
progress in the fabrication and characterization of low-dimensional magnetic
insulators~\cite{Saez2024,Chen2025} will enable experimental realization and testing of
these predictions in the near term.

\sectioncustom{Acknowledgements} A.S.N. and R.E.T. acknowledge funding from Fondecyt Regular 1230515 and 1230747, respectively. This work was funded by  ANID CEDENNA CIA 250002 

\bibliography{memristors}

\end{document}